\documentclass[%
 reprint,
 amsmath,amssymb,
 aps,superscriptaddress,
]{revtex4-1}

\usepackage{graphicx}% Include figure files
\usepackage{braket}
\usepackage{dcolumn}% Align table columns on decimal point
\usepackage{bm}% bold math
\begin{document}

%\preprint{APS/123-QED}

\title{Time-domain stability of parametric synchronization in a spin-torque nano-oscillator based
on a magnetic tunnel junction}% Force line breaks with \\
%\thanks{A footnote to the article title}%

\author{Raghav Sharma}
 \affiliation 
{Department of Physics, Indian Institute of Technology, Hauz Khas, New Delhi-110016, India}%Lines break automatically or can be forced with \\
%\email{phz128316@physics.iitd.ac.in}
%\email{sharmaraghav66@yahoo.com}
\author{Naveen Sisodia}
 \affiliation 
{Department of Physics, Indian Institute of Technology, Hauz Khas, New Delhi-110016, India}%Lines break automatically or can be forced with \\
 
\author{Philipp D\"urrenfeld}%
 %\email{Second.Author@institution.edu.}
\affiliation{School of Electronic Science and Engineering, Nanjing University, 210093 Nanjing, China}
\affiliation{
Department of Physics, University of Gothenburg, 412 96 Gothenburg, Sweden%\\This line break forced with \textbackslash\textbackslash
}%
\author{Johan \AA kerman}%
 %\email{Second.Author@institution.edu.}
\affiliation{
Department of Physics, University of Gothenburg, 412 96 Gothenburg, Sweden%\\This line break forced with \textbackslash\textbackslash
}%
\affiliation{Materials and Nanophysics, School of Engineering Sciences, KTH-Royal Institute of Technology, Electrum 229, 164 40 Kista, Sweden}

\author{Pranaba Kishor Muduli}%
 %\email{Second.Author@institution.edu.}
\affiliation{Department of Physics, Indian Institute of Technology, Hauz Khas, New Delhi-110016, India}%\\This line break forced with \textbackslash\textbackslash
%

% \author{Raghav Sharma}
%  \affiliation{Department of Physics, Indian Institute of Technology, Hauz Khas, New Delhi-110016, India}%Lines break automatically or can be forced with \\
% \author{Naveen Sisodia}%
% \affiliation{Department of Physics, Indian Institute of Technology, Hauz Khas, New Delhi-110016, India}

% \author{{Philipp D\"urrenfeld}}
% \affiliation{School of Electronic Science and Engineering, Nanjing University, 210093 Nanjing, China}
% \affiliation{
% Department of Physics, University of Gothenburg, 412 96 Gothenburg, Sweden%\\This line break forced with \textbackslash\textbackslash
% }%

% \author{Johan \AA kerman}%
%  %\email{Second.Author@institution.edu.}
% \affiliation{
% Department of Physics, University of Gothenburg, 412 96 Gothenburg, Sweden%\\This line break forced with \textbackslash\textbackslash
% }%
% \affiliation{Materials and Nanophysics, School of Engineering Sciences, KTH-Royal Institute of Technology, Electrum 229, 164 40 Kista, Sweden}

% \author{Pranaba Kishor Muduli}%
%  \email{muduli@physics.iitd.ac.in}
% \affiliation{Department of Physics, Indian Institute of Technology, Hauz Khas, New Delhi-110016, India}%\\This

\date{\today}% It is always \today, today,
             %  but any date may be explicitly specified

\begin{abstract}
We report on time-domain stability of the parametric synchronization in a spin-torque nano-oscillator (STNO)
based on a magnetic tunnel junction.  Time-domain measurements of the instantaneous frequency ($f_\textrm{i}$) of a parametrically synchronized STNO show random short-term unlocking of the STNO signal for low injected radio-frequency (RF) power, which cannot be revealed in time-averaged frequency domain measurements. Macrospin simulations reproduce the experimental results and reveal that the random unlocking during synchronization is driven by thermal fluctuations. We show that by using a high injected RF power, random unlocking of the STNO can be avoided. However, a perfect synchronization characterized by complete suppression of phase noise, so-called phase noise squeezing, can be obtained only at a significantly higher RF power. Our macrospin simulations suggest that a lower temperature and a higher positive ratio of the field-like torque to the spin transfer torque reduce the threshold RF power required for phase noise squeezing under parametric synchronization. 
\end{abstract}

% PACS, the Physics and Astronomy
                             % Classification Scheme.
%\keywords{Suggested keywords}%Use showkeys class option if keyword
                              %display desired
\maketitle

%\tableofcontents

\section{\label{sec:level1}Introduction}

Spin-torque nano-oscillators (STNOs)~\cite{kiselev2003nt,deac2008np,vincent2009ieeejssc} are nano-sized microwave devices that can generate a signal in the GHz range by steady magnetization precession of the free magnetic layer. The precession is caused by spin transfer torque~\cite{slonczewski1996jmmm,berger1996prb,ralph2008jmmm}, which is the transfer of angular momentum by a spin-polarized current. The proposed applications of STNO devices in communication systems are supported by the advantageous properties of STNOs, such as a very wide frequency tuning range~\cite{bonetti2010prl,bonetti2009apl,rippard2004prb,muduli2011jap}, high modulation rates~\cite{pufall2005apl,manfrini2009apl,muduli2010prb,muduli2011if,muduli2011ieeem,pogoryelov2011apldc,pogoryelov2011apl}, sub-micron footprints~\cite{vincent2009ieeejssc}, and compatibility with standard complementary metal-oxide semiconductor processes~\cite{engel2005ieeemag,akerman2005sc}. However, practical applications of these devices are hindered by a large linewidth due to high phase noise and by a low output power. A solution to these limitations is the synchronization of multiple oscillators~\cite{mancoff2005nt,kaka2005nt,ruotolo2009phase,sani2013mutually,houshang2016spin,kendziorczyk2016mutual,awad2016long}.

In order to understand mutual synchronization of STNO arrays, it is essential to first understand the synchronization of an STNO to an external microwave signal, also known as injection locking~\cite{rippard2005injection}. While the time-domain stability of STNO injection locking 
is yet to be explored, the impact of thermal fluctuations on injection locking has been reported in some studies~\cite{georges2008coupling,d2010micromagnetic}. Georges \textit{et al.} first discussed the effect of thermal Gaussian noise in spin-valve-based STNOs experimentally and reported a weakening of the synchronization due to thermal Qfluctuations~\cite{georges2008coupling}. However, in another study based on micromagnetic simulations in spin-valve-based STNOs~\cite{d2010micromagnetic}, it was reported that synchronized oscillations should be quite stable with respect to thermal fluctuations, though secondary peaks can appear in the power spectral density. 
In both these studies, the injection locking was done at the fundamental frequency, \textit{i.e.}, at $f_\textrm{RF}\sim f_\textrm{0}$, where $f_\textrm{RF}$ and $f_\textrm{0}$ are the injected radio-frequency (RF) and the STNO frequency, respectively. 

Recently, so-called parametric synchronization of STNOs~\cite{urazhdin2010prl,martin2011parametric,dussaux2011phase,hamadeh2014perfect,durrenfeld2014apl,durrenfeld2014modulation}, where $f_\textrm{RF}$ is about 2$f_\textrm{0}$, was demonstrated experimentally. Parametric synchronization offers several advantages over injection locking done at the fundamental frequency. First, it allows accurate determination of non-linear parameters of STNOs as it favors one single excited mode~\cite{urazhdin2010prl,martin2011parametric,durrenfeld2014apl}.  Secondly, parametric synchronization can also be used to get an enhanced spin torque diode effect at a higher frequency~\cite{tiwari2016enhancement}.  In the context of this work, parametric synchronization greatly simplifies the measurement in time-domain. For the case of synchronization at the fundamental frequency, the RF amplifier can get saturated due to the external RF signal, making it impossible to measure the STNO signal. In case of parametric synchronization, the STNO frequency and input RF are far apart. Hence, by selecting the correct frequency range of the amplifier, its saturation can be avoided. Recently, Demidov \textit{et al.}~\cite{demidov2014synchronization} reported parametric synchronization of spin-Hall nano-oscillators (SHNOs) and showed the threshold nature of synchronization due to the presence of thermal noise.

In this work, we report on time-domain measurements of the parametric synchronization in a magnetic tunnel junction (MTJ) based STNO. We observe a random short-term unlocking of the STNO signal at low RF power. By performing temperature-dependent macrospin simulations, we show that the thermal fluctuations are responsible for the random unlocking of the STNO signal at low RF power and that the random unlocking can be completely removed using a high RF power. However, even higher RF power is required for phase noise squeezing, which produces a signal with minimum linewidth. The term phase noise squeezing describes the condition where the 1/$f^{2}$ phase noise is thoroughly suppressed and was recently reported by Lebrun \textit{et al.} in vortex-based STNOs~\cite{Lebrun2015prl}. 
Our macrospin simulations further suggest that a lower temperature and a higher positive ratio of the field-like torque to the spin transfer torque can reduce the threshold RF power required for phase noise squeezing under parametric synchronization.

\section{\label{sec:method}Methods}

\subsection{\label{sec:Experiment}Experiment}

\begin{figure}
     \includegraphics[width=250pt]{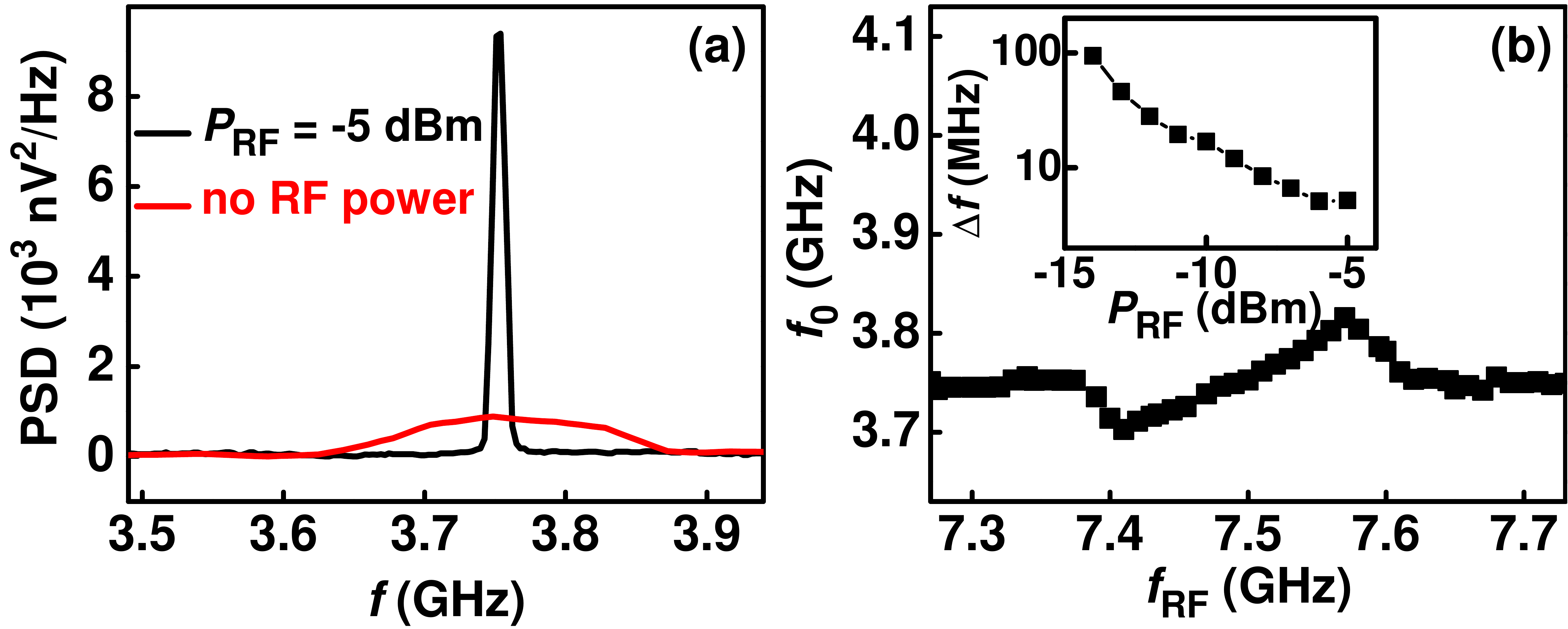}
     
     \caption{\label{fig:fig1} (a) Example spectrum analyzer frequency spectra measured at $I_\textrm{dc}=6$~mA, $H=280$~Oe and $\varphi=196^\circ$. The synchronized signal is measured at $P_\textrm{RF}=-5$~dBm, $f_\textrm{RF}=7.5$~GHz. (b) STNO frequency ($f_\textrm{0}$) vs. RF ($f_\textrm{RF}$) response at $P_\textrm{RF}=-5$~dBm and $I_\textrm{dc}=6$~mA . Inset of Fig. 1(b): Linewidth as a function of $P_\textrm{RF}$ at $f_\textrm{RF}=7.5$~GHz.}    
 \end{figure}

The MTJ-based STNO device under study comprises the following multilayered stack: IrMn(5)/ CoFe(2.1)/ Ru(0.81)/ CoFe(1)/ CoFeB(1.5)/ MgO(1)/ CoFeB(3.5) (thickness values in nm), where the pinned layer is represented by the bottom CoFe layer, the reference layer (RL) is represented by the CoFe/CoFeB layers and the top in-plane magnetized CoFeB layer works as the free layer (FL). We discuss the results from a circular device with an approximate diameter of 180~nm, a resistance-area product of 1.5~$\Omega$($\mu$m)$^{2}$, and a tunneling magnetoresistance of 72\%. We take the direction of the RL magnetization at equilibrium to be $\varphi$ = 0$^\circ$, where $\varphi$ is the angle between the free and the reference layer. The measurements were performed at room temperature. The spectrum analyzer data for the device under study shows a single mode at a frequency of 3.75~GHz [Fig.~\ref{fig:fig1}(a)] at an applied external field of 280~Oe and a dc bias current of $I_\textrm{dc}$~=~6~mA, which is slightly above the threshold current ($I_\textrm{th}$) of 5.9~mA. The field angle of $\varphi$~=~196$^\circ$ was used to maximize the power and obtain a single mode~\cite{muduli2012prl}.

An RF signal with $f_\textrm{RF}=7-8$~GHz (\textit{i.e.,} $\pm500$~MHz from 2$f_\textrm{0}=7.5$~GHz) and the power ranging from $P_\textrm{e}=-6$~dBm to 7~dBm was applied using a Rohde \& Schwarz SMB 100A signal generator through a resistive power divider (0-12~GHz). Here, $P_\textrm{e}$ is the uncorrected power, ignoring losses due to the impedance mismatch and the transmission line. However, in the remainder of this paper we will use $P_\textrm{RF}$, which is the corrected power received by the STNO. The corrected power, $P_\textrm{RF}$ is determined from $P_\textrm{e}$ by taking into account the loss from impedance mismatch (about 60\%)~\cite{durrenfeld2014apl} and losses in the power divider (7~dB). The STNO signal was amplified using a low noise amplifier with a +30~dB gain before being recorded by either a spectrum analyzer or a 4-GHz oscilloscope for frequency and time-domain measurements, respectively.  The operating frequency range of the amplifier (10 - 4200~MHz) only amplifies the STNO signal, while the injected signal (7 - 8~GHz) is effectively suppressed. Furthermore, we use a real-time oscilloscope with a bandwidth of 4~GHz, and hence the injected signal does not appear in the measured time traces. 
  
\subsection{\label{sec:sim} Macrospin simulation}

To understand the origin of the unstable synchronization, we have performed temperature-dependent macrospin simulations~\cite{brown1963thermal,Xiao2005}. For this purpose, we solve the Landau-Lifshitz-Gilbert-Slonczewski (LLGS) equation in macrospin approximation~\cite{zhu2012voltage,zeng2015spin}:

\begin{equation} \label{eq:1}
	\begin{gathered}
\frac{d\hat{m}}{dt}=-\gamma(\hat{m}\times \vec{H}_{eff}) + \alpha(\hat{m}\times \frac{d\hat{m}}{dt}) \\
- \gamma\frac{J\hbar P}{2eM_{s}d(1+P^2\cos\phi)}[\hat{m}\times(\hat{m}\times\hat{e}_{p})
+ b_{f}(\hat{m}\times\hat{e}_{p})]
	\end{gathered}
\end{equation}

Here, $\hat{m}$ is the normalized magnetic moment, $\gamma$ is the gyromagnetic ratio of the electron, $J$ is the polarized current density, $P$ is the polarization efficiency, $e$ is the electronic charge, and $\hat{e}_p$ is the direction of the fixed layer with $\phi$ as the angle between free and fixed layer. The ratio of field like torque to in-plane torque is represented by $b_{f}$. The saturation magnetization of the free layer, $M_{s}$, and the Gilbert damping constant, $\alpha$, were taken as $10^{6}$~A/m and 0.02, respectively. The fixed layer was assumed to be oriented along the $\hat{x}$ axis having a polarization efficiency of 0.65. The diameter of the simulated circular free layer (240~nm) was different from the actual device (180~nm) to match the threshold current. The $I_\textrm{dc}/I_\textrm{th}$ ratio is maintained $\sim$1 for both the cases of experiment and simulation. The thickness of the free layer, $d$, was taken as 3.5~nm. Note that these parameters are chosen to qualitatively reproduce the experimental results. To solve the LLGS equation, it was first discretized in time-domain and then integrated in small time steps (0.1~ps) using a fourth order Runge-Kutta method for the total time period of 250 $\mu$s. The effective field, $\vec{H}_{eff}$, in Eq.~(\ref{eq:1}) is the sum of external field, demagnetizing field and thermal field:
\begin{equation} \label{eq:2}
\vec{H}_{eff}=\vec{H}_{ext}+\vec{H}_{demag}+\vec{H}_{T}{.}
\end{equation}
The external field, $\vec{H}_{ext}$, is directed in plane, making an angle of $196^\circ$ with the $\hat{x}$ axis:
\begin{equation} \label{eq:3}
\vec{H}_{ext}=|H_x|\hat{x}+|H_y|\hat{y}{.}
\end{equation}
The demagnetizing field, $\vec{H}_{demag}$, which depends on the shape of the ferromagnet and the instantaneous magnetization, $\hat{m}$, is given by
\begin{equation} \label{eq:4}
\vec{H}_{demag}= -M_s(N_x m_x\hat{x}+N_y m_y\hat{y}+N_z m_z\hat{z}){,}
\end{equation}
where $N_x$, $N_y$, and $N_z$ represent the demagnetization factors in $\hat{x}$, $\hat{y}$, and $\hat{z}$ directions, respectively. Their values have been approximated for the case of a thin circular disk where the thickness is much lower than the diameter~\cite{Osborn1945}. The calculated values of $N_x$, $N_y$, and $N_z$ are 0.01125, 0.01125 and 0.9775, respectively.

The contribution of random fluctuations due to finite temperatures is included in the thermal field, $\vec{H}_{T}$. Each Cartesian component of $\vec{H}_{T}$ satisfies the Brown's approximation~\cite{brown1963thermal,Xiao2005}:
\begin{equation} \label{eq:5}
\braket{H_{\rm T}^i(t)H_{\rm T}^j(t)}=\frac{2k_{\text{B}}T\alpha}{\gamma V \mu_{0} M_{s}} \delta_{ij}\delta(t-t'){,}
\end{equation}
where the variables $i$ and $j$ refer to components of the time-varying thermal field in different directions. $k_{\rm B}$ is the Boltzmann constant, $T$ is the temperature, $V$ is the volume of the free layer and $\mu_{0}$ the magnetic permeability of free space. The variance of the thermal field according to above Eq.~\eqref{eq:5} ensures that the system produces a Boltzmann distribution of energies at temperature $T$.

The simulated STNO signal is found to have a single mode signal at 4.75~GHz for $I_\textrm{dc}$ = 6~mA (not shown). Since we have not considered a field due to inter layer exchange coupling in our simulations~\cite{muduli2011prb}, such a frequency difference between simulation and experiment is not unexpected.

\begin{figure}[t!]
    \includegraphics[width=255pt]{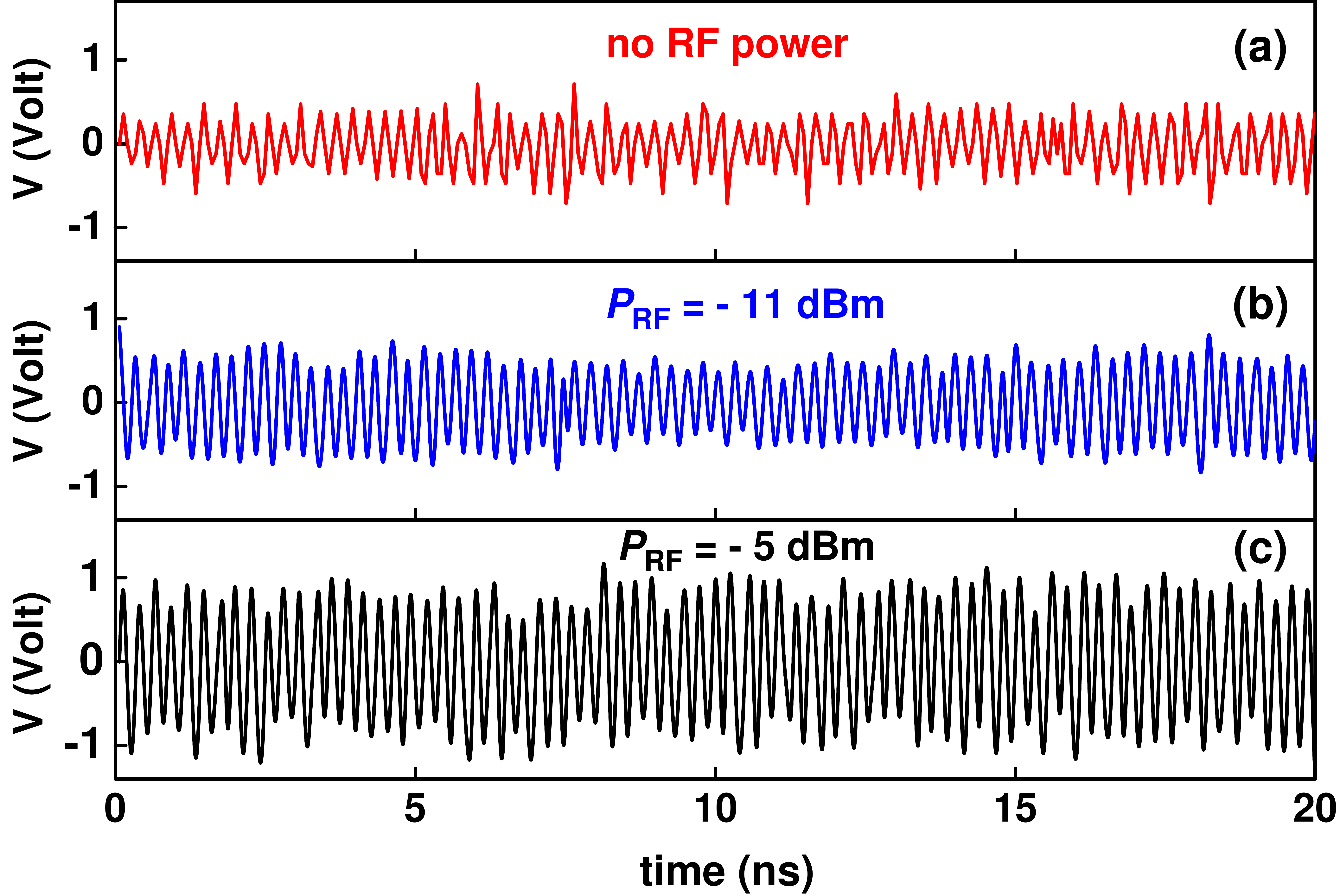}
    \caption{\label{fig:fig2_1} Example time traces measured at $I_\textrm{dc}=6$~mA, $H=280$~Oe and $\varphi=196^\circ$ for (a) free running STNO (no RF injection), (b) $P_\textrm{RF}=-11$~dBm and $f_\textrm{RF}=7.5$~GHz, and (c) $P_\textrm{RF}=-5$~dBm and $f_\textrm{RF}=7.5$~GHz. }
\end{figure}

\section{\label{sec:res}Results and discussion}
\subsection{\label{sec:tdst} Time-domain stability of parametric synchronization}
\begin{figure}
    \includegraphics[width=250pt]{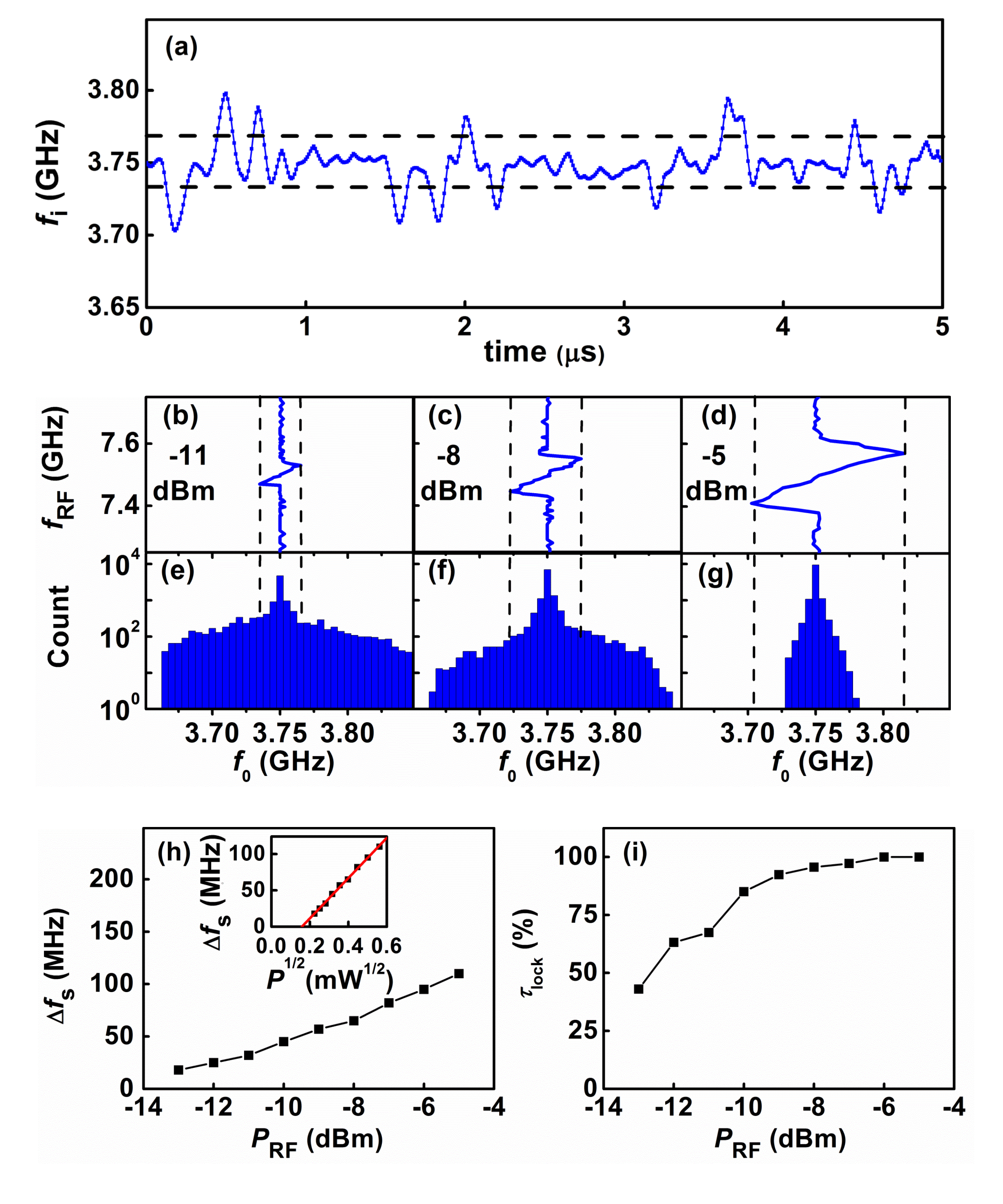}
    \caption{\label{fig:fig2} (a) Measured instantaneous frequency $f_\textrm{i}$ at $P_\textrm{RF}=-11$~dBm. The two horizontal lines show the synchronization interval, which is $\sim$30 MHz for this case. (b-d) Frequency domain measurements of the frequency of the STNO, $f_\textrm{0}$ vs. $f_\textrm{RF}$ at $P_\textrm{RF}=-11, -8$ and -5~dBm, respectively. (e-g) Histograms showing the distribution of $f_\textrm{i}$ for a total number of bins of 12500 at $P_\textrm{RF}$ = -11, -8 and -5~dBm, respectively. The resolution on the \textit{x}-axis is 5~MHz. (h) The Synchronization interval, $\Delta f_\textrm{s}$ vs. $P_\textrm{RF}$  Inset : Synchronization interval vs. $P^{1/2}$. Solid line is a linear fit. (i) $\tau_\textrm{lock}$ vs. $P_\textrm{RF}$.}
\end{figure}
A comparison between the spectrum analyzer power spectral density of the free-running and the synchronized signal at $P_\textrm{RF}=-5$~dBm is shown in Fig.~\ref{fig:fig1}(a). Figure~\ref{fig:fig1}(b) shows the frequency of the STNO as a function of $f_\textrm{RF}$ for $P_\textrm{RF}=-5$~dBm at a dc bias current of $I_\textrm{dc}=6$~mA. The STNO frequency is locked in the range of $f_\textrm{RF, min}=7.4$~GHz to $f_\textrm{RF, max}=7.6$~GHz, resulting in a synchronization interval of $\Delta f_\textrm{s}=(f_\textrm{RF, max}-f_\textrm{RF, min})/2\approx$107~MHz. The linewidth $\Delta f$ is found to significantly decrease from the free-running value of 115~MHz to 5~MHz at $P_\textrm{RF}=-5$~dBm [inset of Fig.\ref{fig:fig1}(b)], while the peak power increases from 840~nV$^{2}$/Hz to 9553~nV$^{2}$/Hz. In this work, the reduction in linewidth is by a factor of 23 during parametric synchronization. A further reduction of linewidth would require applying a higher RF power. However, the maximum RF power that can be applied to our MTJ without the breakdown of the barrier is -5~dBm. Contrary to our last work~\cite{durrenfeld2014apl}, where the linewidth of the free-running STNO was 10-20~MHz, the free-running linewidth in this work is 115~MHz due to a different operating condition, namely, a lower $f_\textrm{0}$. This condition is chosen in order to be able to study the stability of the parametric synchronization in the time-domain. As we will show later, the relative reduction in linewidth in this work is less due to this large initial free-running linewidth.  
\begin{figure}
    \includegraphics[width=252pt]{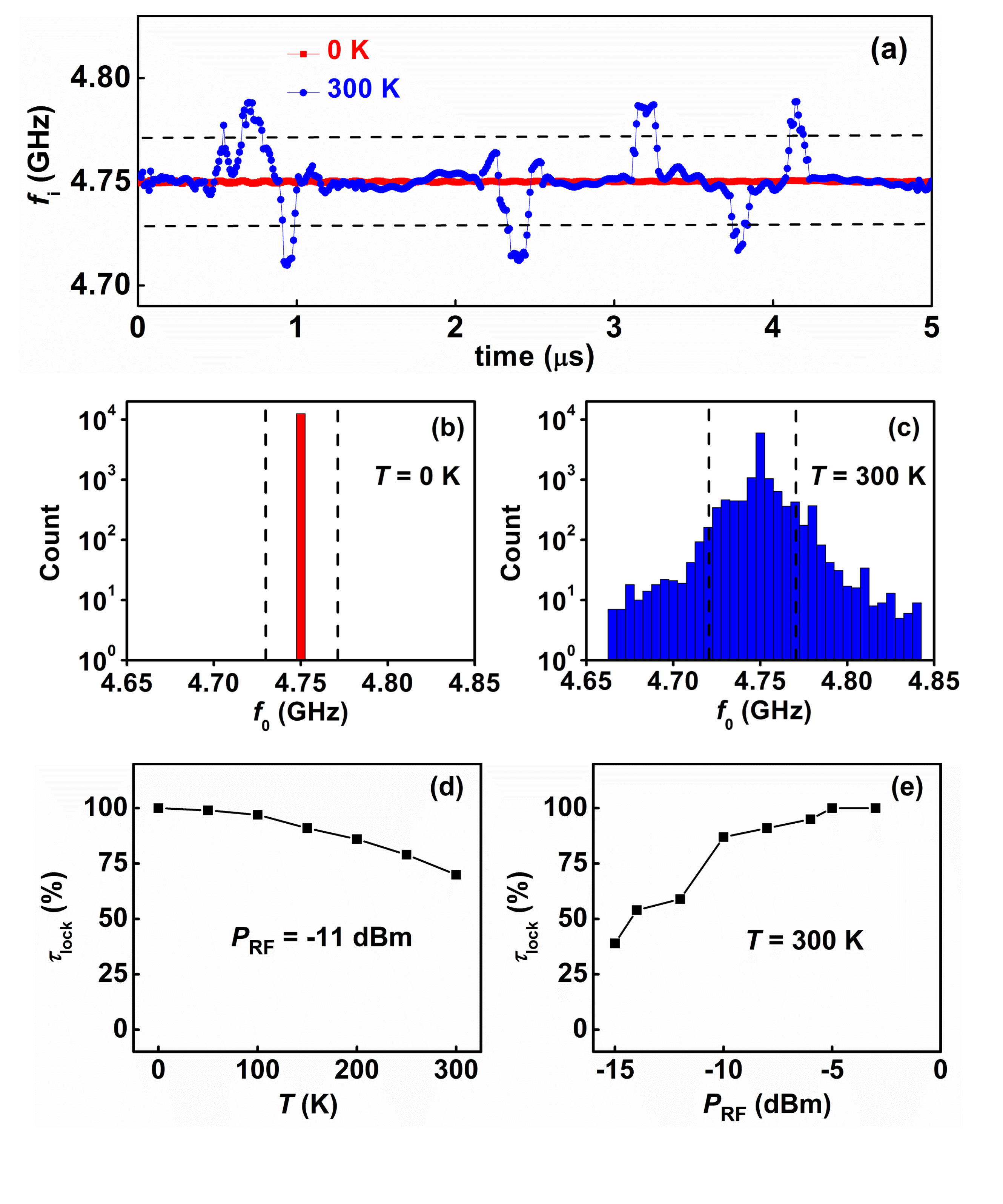}
    \caption{\label{fig:macrospinfi} (a) Simulated instantaneous frequency for the locked signal at $T=0$~K and 300~K at $f_\textrm{RF}=9.5$~GHz. (b,c) Histograms showing the distribution of the instantaneous STNO frequency at $T= 0$~K and 300~K, respectively. The resolution on the \textit{x}-axis is 5~MHz. (d) Simulated $\tau_\textrm{lock}$ vs. temperature at $f_\textrm{RF}=9.5$~GHz and $P_\textrm{RF}=-11$~dBm. (e) $\tau_\textrm{lock}$ vs $P_\textrm{RF}$ at $f_\textrm{RF}=9.5$~GHz.}   
\end{figure}

In order to study the time-domain stability of the synchronization, 250~$\mu$s long time traces were captured in the oscilloscope. Figures~\ref{fig:fig2_1}(a)-2(c) show short sections (20~ns) of the measured time-domain traces for the free running and the synchronized signal at $P_\textrm{RF}=-11$~dBm and $-5$~dBm. The time traces show improvement in signal quality with increase in RF power. In order to better understand the synchronization process, the instantaneous frequency ($f_\textrm{i}$) of the time-trace is determined by taking the derivative of the instantaneous phase, which is calculated using the Hilbert transform method for each 10~ns long segment of the STNO signal~\cite{bianchini2010apl}.
Figure~\ref{fig:fig2}(a) shows the derived $f_\textrm{i}$ vs. time at $P_\textrm{RF}=-11$~dBm for a 5~$\mu$s long section of the time trace. The two black dotted lines are defined by $f_\textrm{RF, max}/2$ and $f_\textrm{RF, min}/2$. As can be seen, the fluctuations in $f_\textrm{i}$ are larger than the synchronization interval with random short-term unlocking. Figures~\ref{fig:fig2}(b)-3(g) compare the frequency and time-domain measurements. The frequency domain measurement of the response of the STNO frequency with $f_\textrm{RF}$ are shown in Figs.~\ref{fig:fig2}(b)-3(d) for $P_\textrm{RF}= -11$, -8 and -5~dBm, respectively. The corresponding  histogram of $f_\textrm{i}$ from the time-domain measurements are shown in Fig.~\ref{fig:fig2}(e)-3(g). The vertical lines indicate the synchronization interval. At $P_\textrm{RF}=-11$~dBm the histogram extends beyond the synchronization interval, while at $P_\textrm{RF}=-5$~dBm the histogram lies completely within the synchronization interval. The synchronization interval increases with RF power as shown in Fig.~\ref{fig:fig2}(h). The inset of Fig.~\ref{fig:fig2}(h) shows the plot of the synchronization interval vs. $P^{1/2}$, where $P$ is the corrected power in milliwatts. The behavior is linear and it exhibits a threshold in agreement with the recent work on synchronization of SHNOs,~\cite{demidov2014synchronization} where the non-zero threshold was explained by thermal fluctuations. 

For quantifying the effective locking time, we define a quantity $\tau_\textrm{lock}$ as the time for which $f_\textrm{i}$ lies within the boundaries of $f_\textrm{RF, max}/2$ and $f_\textrm{RF, min}/2$. We express $\tau_\textrm{lock}$ in percentage by dividing with the total measured time of 250~$\mu$s. We found $\tau_\textrm{lock}$ to be only 43\% at $P_\textrm{RF}=-13$~dBm and 63\% at $P_\textrm{RF} = -11$~dBm indicating frequent unlocking/locking events. However, $\tau_\textrm{lock}$ is improved with the increase of RF power and reaches 100\% at $P_\textrm{RF}=-6$~dBm as shown in Fig.~\ref{fig:fig2}(i). The time-domain instability of synchronization, characterized by $\tau_\textrm{lock}\neq 100\%$ is undesirable for any practical applications, especially for injection-locked phase locked loops (PLLs), which are one of the most important building blocks of RF transceivers. Thus our results show that the frequency domain data, which indicates a locking already at a low power of $P_\textrm{RF}=-13$~dBm, can be misleading, while $\tau_\textrm{lock}$ reached 100\% only at $P_\textrm{RF}=-6$~dBm.  However, we will show in Sec.\ref{sec:phasenoise} that $\tau_\textrm{lock}=100$\% does not imply a “perfect synchronization”, which is defined by the phase noise squeezing.

In order to investigate the origin of the above unstable synchronization, we perform finite temperature based macrospin simulations. Figure~\ref{fig:macrospinfi}(a) shows the simulated instantaneous frequency calculated from the Hilbert transform method at $P_\textrm{RF}=-11$~dBm, a condition close to the experimental measurements.   
The synchronized state is perfect at $T=0$~K with negligible fluctuations in $f_\textrm{i}$ throughout the total simulated time of 250~$\mu$s. However, at 300~K, $f_\textrm{i}$ shows increased random fluctuations that go beyond the synchronization interval. Figure~\ref{fig:macrospinfi}(b) and 4(c) show the histograms of the instantaneous frequency for $T = 0$~K and $T = 300$~K, respectively. The dotted lines show the synchronization interval, which is 46 MHz for $T=0$~K and 42 MHz for $T=300$~K. These figures demonstrate random unlocking of STNO at 300~K. The temperature dependence $\tau_\textrm{lock}$ is shown in  Fig.~\ref{fig:macrospinfi}(d) which shows a reduction of $\tau_\textrm{lock}$  with temperature. $\tau_\textrm{lock}$  is reduced to about 70\% at $T = 300$~K from 100\% at $T= 0$~K. These simulations suggest that the random fluctuations in $f_\textrm{i}$ in the experiment are caused by thermal fluctuations, which cause the STNO frequency to come out of the synchronization interval.  However, it is possible to achieve perfect synchronization by applying a higher RF power which overcomes the thermal fluctuations. As shown in Fig.~\ref{fig:macrospinfi}(e), for  $P_\textrm{RF} = -5$~dBm, the synchronization is stable and $\tau_\textrm{lock}$ reaches 100\% at $T=300$~K. We also examined the phase difference between the injected signal and the STNO for the case of $f_\textrm{RF}=2f_\textrm{0}$ and found that random phase slips occur at low RF power of $P_\textrm{RF}=-11$~dBm . As the RF power is increased and a $\tau_\textrm{lock}$ of 100\% is achieved,  we only observe phase slips of $\pi$, which, according to Ref.~\cite{Lebrun2015prl}, is a signature of phase locking with an injected signal at $f_\textrm{RF}=2f_\textrm{0}$.

\subsection{\label{sec:phasenoise} Phase noise under parametric synchronization}

\begin{figure}[t!]
    \includegraphics[width=255pt]{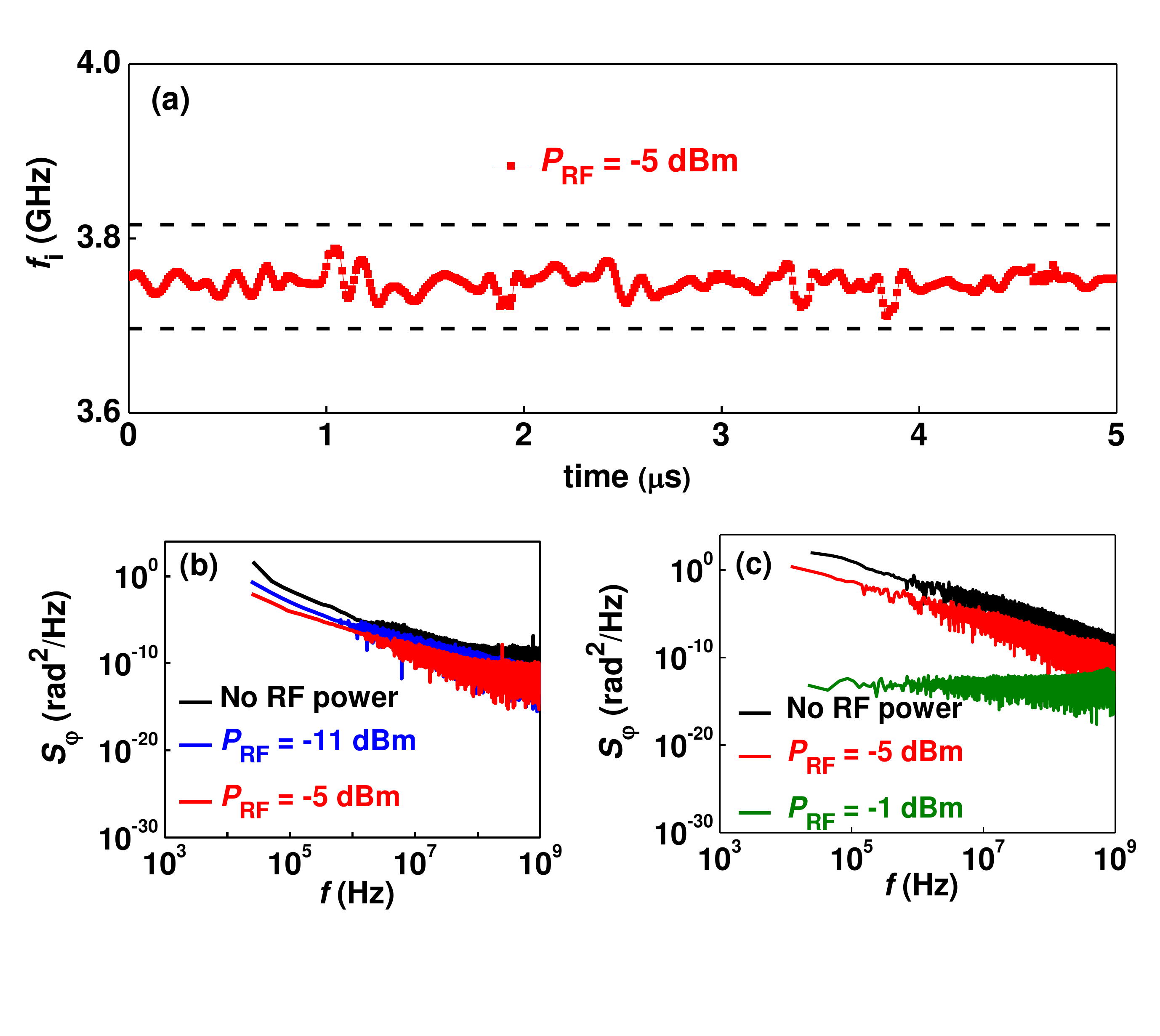}
    \caption{\label{fig:fig4} (a) Experimentally measured instantaneous frequency $f_\textrm{i}$ of the synchronized signal at $P_\textrm{RF} = -5$~dBm. (b) Experimentally measured phase noise at $f_\textrm{RF}=7.5$~GHz for $I_\textrm{dc}=6$~mA with varying RF power $P_\textrm{RF}$. (c) Simulated phase noise at $f_\textrm{RF}=9.5$~GHz for $I_\textrm{dc}=6$~mA with varying RF power $P_\textrm{RF}$.}
\end{figure} 

Our experimental results show that at $P_\textrm{RF}=-5$~dBm, we achieve $\tau_\textrm{lock}=100$\%, yet the linewidth of the STNO does not reach that of the injected RF signal at this power. In order to obtain a better understanding of this behavior, we studied the phase noise as a function of injected RF power in both experiment and simulations. Figure~\ref{fig:fig4}(a) shows the experimental $f_\textrm{i}$ at the highest RF power of $P_\textrm{RF}= -5$~dBm. The fluctuations in $f_\textrm{i}$ are reduced considerably at this power and all fluctuations are within the synchronization interval of 107~MHz shown by the dotted lines. In Fig.~\ref{fig:fig4}(b) we display the experimental phase noise at RF powers of $P_\textrm{RF}=-5$~dBm 
and -11~dBm 
along with the free running case of no RF power  
The phase noise is estimated by utilizing the zero crossing method~\cite{keller2010nonwhite,raghavapl2014}. 
Clearly, the more stable locked signal yields less $1/f^{2}$ noise due to the reduced frequency fluctuations and an increased $\tau_\textrm{lock}$ around the central frequency of $f_\textrm{0}=3.75$~GHz. However, experimentally the phase noise is not completely reduced or squeezed (so called phase noise squeezing) even after $\tau_\textrm{lock}$ reaches 100\% at $P_\textrm{RF}=-5$~dBm.

The simulations performed for conditions similar to experiments, reveal that a much higher $P_\textrm{RF} = -1$~dBm,  
is needed for a complete phase noise squeezing, as shown in Fig.~\ref{fig:fig4}(c), although $\tau_\textrm{lock}$ reaches 100\% already at $P_\textrm{RF}=-5$~dBm. This is similar to our previous experimental observation, where at $P_\textrm{RF}=-5$~dBm, $\tau_\textrm{lock}$ is 100\%, yet without phase noise squeezing. At the condition of phase noise squeezing, we observe no phase slips and phase difference between the injected signal and STNO is constant with time. However, a considerable higher RF power is needed for the phase noise squeezing. We define $P_\textrm{RF}^{sq}$ as the RF power required for phase noise squeezing. For this case $P_\textrm{RF}^{sq}=-1$~dBm, which relates to a ratio of RF current to dc bias current of $I_\textrm{RF}/I_\textrm{dc}$~=~0.5. This is a high RF power in comparison to the maximum possible experimental power $P_\textrm{RF}=-5$~dBm for our STNO devices at $I_\textrm{dc}=6$~mA. This value is also larger compared to previously reported values of RF power required for phase noise squeezing in vortex based STNOs~\cite{Lebrun2015prl}.

\begin{figure}[t!]
\includegraphics[width=250pt]{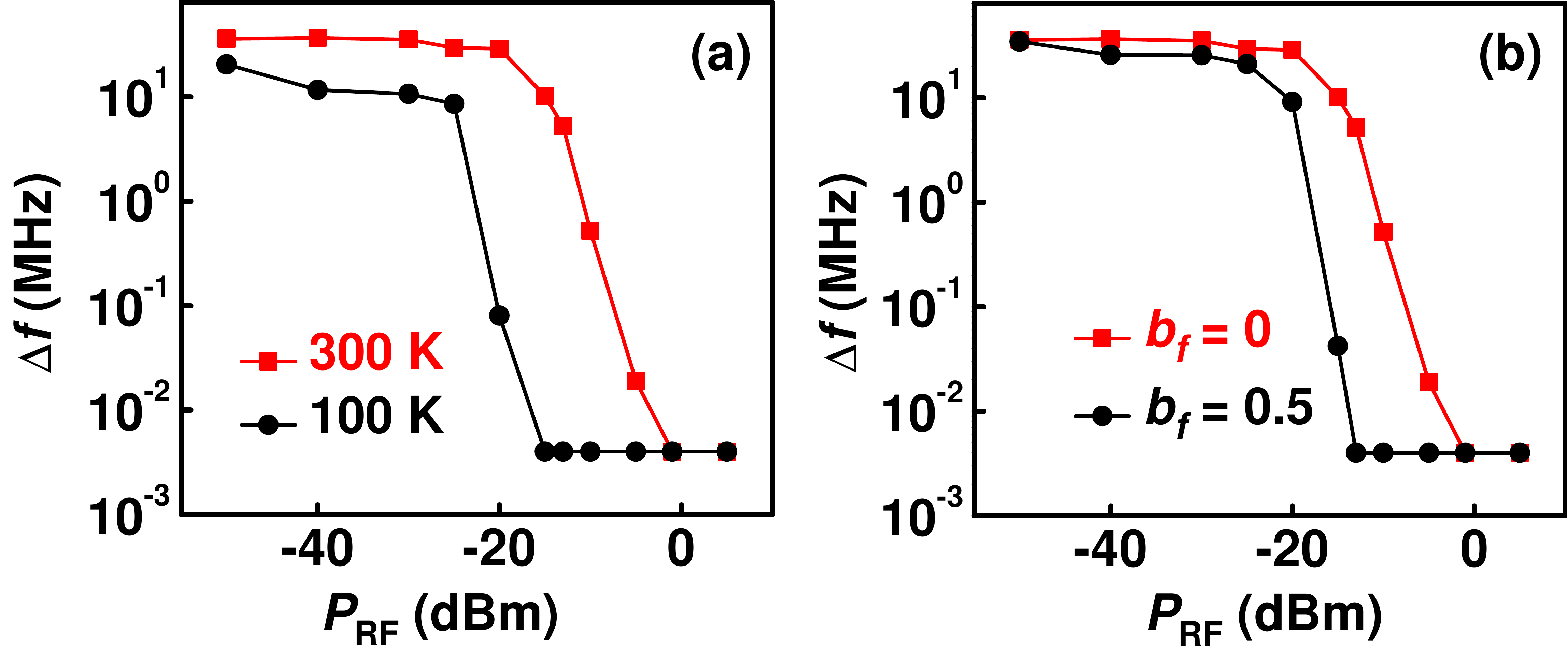}
    \caption{\label{fig:fig5}  Simulated linewidth  vs. RF power at $f_{RF}=2f_\textrm{0}$ and $I_\textrm{dc}=6$~mA (a) for two selected temperatures at $T=100$~K and 300~K and (b) for two cases of field-like torque $b_{f}=0$ and $b_{f}=0.5$ at $T=300$~K.}
\end{figure}

In order to reduce the RF power required for phase noise squeezing, we studied the effect of temperature and field-like torque on $P_\textrm{RF}^{sq}$ by macrospin simulations. We simulated 250~$\mu$s long time traces, for which phase noise squeezing will lead to a minimum linewidth of $\Delta f=4$~kHz due to the finite length of the time trace. Figure~\ref{fig:fig5}(a) shows the variation of linewidth with RF power at the temperatures of 100~K and 300~K. At 100~K, the threshold RF power required for phase noise squeezing, \textit{i.e.}, when the linewidth reaches 4~kHz, is significantly lower compared to 300~K. Thus, $P_\textrm{RF}^{sq}$ strongly depends on temperature. In fact, for the case of $I_\textrm{dc}=6$~mA, $P_\textrm{RF}^{sq}$ is reduced from -1~dBm at 300~K to -15~dBm at 100~K. We also note that the free running linewidth at 300~K is larger and this also leads to a large $P_\textrm{RF}^{sq}$. This explains why we achieve a lower reduction in linewidth in this work [see inset of Fig. 1 (b)] as compared to our previous work~\cite{durrenfeld2014apl} for the same level of injected RF power. 
Our detailed current-dependent room-temperature simulations show that $P_\textrm{RF}^{sq}$  reduces to -7~dBm for a higher $I_\textrm{dc}$ of 7~mA (not shown) compared to -1~dBm at $I_\textrm{dc}=6$~mA. This is expected, since a higher dc current leads to a larger spin torque and hence less influence of the thermal fluctuations.   

We also studied the effect of the ratio of field-like torque to spin transfer torque as shown in Fig.~\ref{fig:fig5}(b). The figure shows the variation of linewidth with RF power for $b_{f}=0$  and $b_{f}=0.5$. We note that the high value of $b_{f}=0.5$ is consistent with the previous reports where field like torque is reported to be 40-50\% of the spin transfer torque~\cite{wang2011time,matsumoto2011spin}. Please note that all the previous simulations were shown for $b_{f}=0$. 
We found that $P_\textrm{RF}^{sq}$ is reduced from -1 to -13~dBm  with an increase of $b_{f}$ from 0 to 0.5 at $T=300$~K. Thus, a lower temperature and a higher field-like torque can help in achieving phase noise squeezing at low RF power. We believe that the behavior of $P_\textrm{RF}^{sq}$ with the field-like term is similar to the case of vortex oscillators~\cite{Lebrun2015prl}. It is caused by a reduction of the phase difference between the injected signal and the STNO phase by the field-like torque. However, the reduction in $P_\textrm{RF}^{sq}$  with the field like-term is found only for a positive sign of $b_{f}$ in Eq.~(\ref{eq:1}). A negative sign of $b_{f}$ leads to an increase in  $P_\textrm{RF}^{sq}$ (not shown).

In conclusion, we reported on a time-domain study of parametrically synchronized STNO signals. Experimental analysis shows a time-domain instability of synchronization and random unlocking at low RF power, which improves by increasing the strength of RF power. Macrospin simulations show that the time-domain instability of synchronization is caused by the finite temperature. We show that phase noise squeezing, which is a signature of perfect synchronization, is easier to achieve at lower temperatures and with a larger and positive field-like torque. These results indicate that at finite temperature the threshold RF power required for phase noise squeezing is significantly higher. Hence, for achieving perfect synchronization of multiple STNOs, a low free-running linewidth of STNOs, and less thermal fluctuations are desirable.

\begin{acknowledgments}
Partial financial support by the Department of Science and
Technology (India) under Fast-Track Scheme and the Department
of Electronics and Information Technology (DeitY),
Government of India is gratefully acknowledged. Financial
support from the Swedish Foundation for Strategic Research
(SSF) and the Swedish Research Council (VR) are gratefully
acknowledged. Authors thank the IIT Delhi HPC facility for
computational resources. R.S and N.S acknowledge support
from Ministry of Human Resource Development (MHRD),
India.
\end{acknowledgments}

\end{document}